\documentclass[reprint,amsmath,amssymb,aps,prl]{revtex4-1}

\usepackage{graphicx}
\usepackage{dcolumn}
\usepackage{bm}
\usepackage{hyperref}
\usepackage{xcolor}

\begin{document}

\title{Type-II Ising superconductivity in two-dimensional materials with spin-orbit coupling}

\author{Chong \surname{Wang}$^{1,2,3,4}$}
\author{Biao \surname{Lian}$^{5}$}
\author{Xiaomi \surname{Guo}$^{2,3}$}
\author{Jiahao \surname{Mao}$^{2,3}$}
\author{Zetao \surname{Zhang}$^{2,3}$}
\author{Ding \surname{Zhang}$^{2,3,6}$}
\author{Bing-Lin \surname{Gu}$^{1,2,3}$}
\author{Yong \surname{Xu}$^{2,3,4}$}
\email{yongxu@mail.tsinghua.edu.cn}
\author{Wenhui \surname{Duan}$^{1,2,3}$}

\affiliation{$^{1}$Institute for Advanced Study, Tsinghua University, Beijing 100084, China\\
$^{2}$State Key Laboratory of Low Dimensional Quantum Physics and Department of Physics, Tsinghua University, Beijing, 100084, China\\
$^{3}$Collaborative Innovation Center of Quantum Matter, Tsinghua University, Beijing 100084, China\\
$^{4}$RIKEN Center for Emergent Matter Science (CEMS), Wako, Saitama 351-0198, Japan\\
$^{5}$Princeton Center for Theoretical Science, Princeton University, Princeton, NJ, USA\\
$^{6}$Beijing Academy of Quantum Information Sciences, Beijing 100193, China
}


\begin{abstract}
Centrosymmetric materials with spin-degenerate bands are generally considered to be trivial for spintronics and related physics. In two-dimensional (2D) materials with multiple degenerate orbitals, we find that the spin-orbit coupling can induce spin-orbital locking, generate out-of-plane Zeeman-like fields displaying opposite signs for opposing orbitals, and create novel electronic states insensitive to in-plane magnetic field, which thus enables a new type of Ising superconductivity applicable to centrosymmetric materials. Many candidate materials are identified by high-throughput first-principles calculations. Our work enriches the physics and materials of Ising superconductivity, opening new opportunities for future research of 2D materials.
\end{abstract}

\maketitle

Two-dimensional (2D) materials have attracted enormous research interest due to their extraordinary electronic properties, where the interplay of spin-orbit coupling (SOC) and symmetry provides a fertile ground to discover emerging quantum phenomena, including 2D Ising ferromagnetism, valley-contrasting physics, the quantum spin/anomalous Hall effects, topological magnetoelectric effects, and so on\cite{huang_layer-dependent_2017, gong_discovery_2017,xiao_coupled_2012,hasan2010,qi2011}. Thanks to these outstanding discoveries, the profound role of time reversal symmetry (TRS) and inversion symmetry in the SOC-related physics has been well recognized. By generalizing the concept further to 2D superconductors, where the $U(1)$ symmetry is broken, even more intriguing emergent physics is expected\cite{saito_highly_2017}. One long-sought object of superconductivity research is to find superconductors robust against a magnetic field. The great advantage of 2D superconductors is that orbital effects of in-plane magnetic field are significantly eliminated by quantum confinement\cite{tinkham_introduction_1996}. Then paramagnetic effects will completely suppress superconductivity when the Zeeman energy splitting overcomes the binding energy of Cooper pairs, setting the Pauli paramagnetic limit ($B_p$) as the upper critical field ($B_{c2}$) \cite{clogston_upper_1962, chandrasekhar_note_1962}. This limit, however, might be overcome by SOC that introduces an emergent magnetic field compatible with superconductivity\cite{manchon_new_2015}.  

Early attempts proposed to enhance $B_{c2}$ by introducing SOC effects extrinsically into superconductors, for instance, by spin-orbit scattering\cite{klemm_theory_1975}, which unfortunately is accompanied with complicated and possibly deleterious influence of defects and disorders. In contrast, it is more favorable to incorporate SOC effects through intrinsic ways. Previous studies found that Rashba-type SOC in 2D noncentrosymmetric superconductors can create novel mixed singlet-triplet pairing and leads to a moderate enhancement of $B_{c2}$\cite{gorkov_superconducting_2001, yip_two-dimensional_2002}. Very recently, a new kind of 2D noncentrosymmetric superconductivity, called Ising superconductivity, has been discovered in transition metal dichalcogenides (TMDs)\cite{lu_evidence_2015,saito_superconductivity_2016,zhou_ising_2016,liu_interface-induced_2018,xi_ising_2016,lu_full_2018,barrera_tuning_2018}. In these materials, the SOC generates Zeeman-like spin splittings for electrons near the $K$ and $K'$ valleys. Under the extraordinarily large SOC fields, spins of Cooper pairs are aligned along the out-of-plane direction and effectively do not respond to in-plane magnetic field, resulting in enhanced $B_{c2}$ considerably beyond the Pauli limit. This breakthrough has stimulated great research interests in searching for Ising superconductors alike.

However, previous search of Ising superconductors was limited to noncentrosymmetric materials. The fundamental reason is that a centrosymmetric material with TRS is forbidden by symmetry to have spin-split bands, which does not fit the existing scenario of Ising superconductivity. In this Letter, we propose a new type (dubbed as type-II) of Ising superconductivity that does not involve inversion asymmetry, based on 2D materials with multiple degenerate orbitals. Despite the spin degeneracy, the SOC can lead to a so-called spin-orbital locking\cite{fu_parity-breaking_2015,zhang_spin-orbital_2013} in centrosymmetric materials, displaying opposite Zeeman-like fields for opposing orbitals. This emergent Zeeman field strongly polarizes electron spins out-of-plane, which creates novel electronic states insensitive to in-plane magnetic field and thus enables Ising superconducting pairing. By high-throughput first-principles calculations, we predict a large number of candidate 2D materials, including their monolayers and thin films, which might host type-II Ising superconductivity. The finding might open new directions for fundamental research and practical applications of 2D materials.

\begin{figure}
\includegraphics[width=1.0\columnwidth]{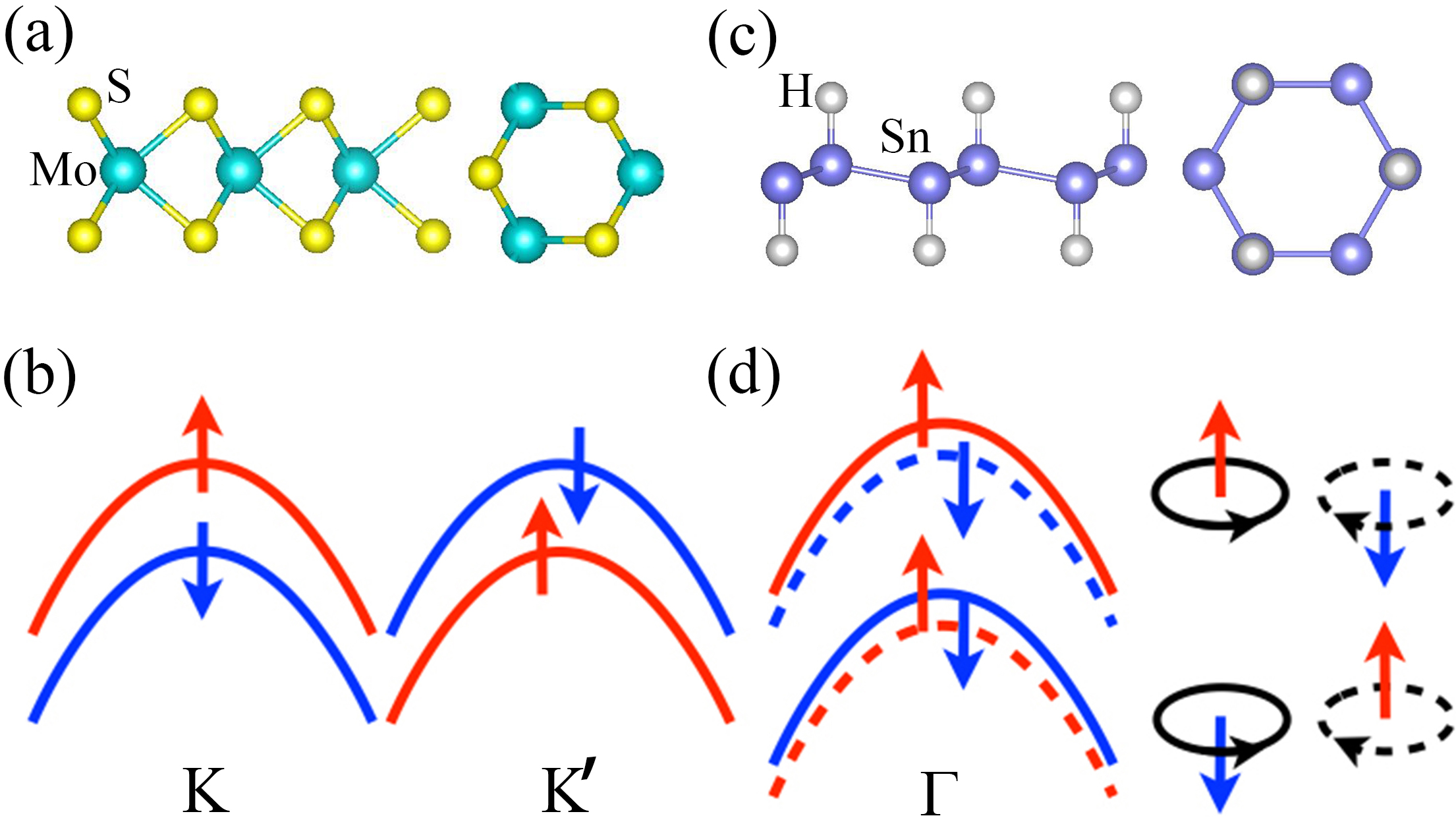}
\caption{Schematic picture of Ising superconductivity. (a) presents monolayer MoS$_2$ with type-I Ising superconductivity sketched in (b). Near $K$ and $K'$, spin up (red) and spin down (blue) split in energy by a Zeeman-like field, which is opposite for the two valleys. (c) presents monolayer SnH with type II-Ising superconductivity sketched in (d). Near $\Gamma$, spin up and spin down are locked to the two opposing orbitals (denoted by solid and dashed lines, respectively). They spit in energy for each orbital by a Zeeman-like field, whose sign is opposite for opposing orbitals. {The dashed and solid lines are degenerate in energy.}
\label{fig1}}
\end{figure}

The interplay of SOC and symmetry plays a key role in defining Ising superconductivity, as illustrated in Fig. \ref{fig1}. For the previous type (named type-I) that appears in the 2H phase of TMDs (like MoS$_2$), inversion symmetry has to be broken. The SOC thus can result in Zeeman-like spin splitting near the $K$ and $K'$ valleys. Moreover, the Zeeman-like splitting is opposite for the two valleys, so as to preserve TRS. In the absence of external perturbation, the Zeeman-like field is along the out-of-plane ($z$) direction, benefiting from the mirror symmetry $M_z$ of materials. Such kind of SOC effects are described by an effective Hamiltonian $H^\textrm{I}_\textrm{SOC} = \beta_\textrm{SO} s_z \sigma_z$, where $\beta_\textrm{SO}$ denotes the SOC strength, $s_z = \pm$ labels the $K$($K'$) valleys, and $\sigma_z =\pm$ labels spin up(down). Electron spins are aligned out-of-plane by the effective Zeeman field $\beta_\textrm{SO}s_z$ and become insensitive to in-plane magnetic field. Type-I Ising superconducting pairing of opposite spins and valleys is thus enabled. 

In contrast, type-II Ising superconductivity to be proposed is applicable to both centrosymmetric and noncentrosymmetric 2D materials with multiple degenerate orbitals. We will first study centrosymmetric materials and discuss the influence of inversion symmetry breaking later. The orbital degeneracy can be ensured by $n$-fold rotational symmetry $C_{nz}$ ($n=3,4,6$) and TRS, or by other symmetries. To demonstrate the concept, we study an exemplary material--hydrogen-saturated stanene (SnH) [Fig. \ref{fig1}(c)] that has the $D_{3d}$ space group and degenerate $p_{x,y}$ orbitals in the top two valence bands. This material system displays interesting topological quantum physics\cite{xu_large-gap_2013,deng_epitaxial_2018,zhu_epitaxial_2015,molle_buckled_2017}, and few-layer stanene was recently discovered to be a 2D superconductor\cite{liao_superconductivity_2018}.

At $C_{3z}$ invariant momenta of the Brillouin zone (e.g. $\Gamma$), Bloch states can be labeled by $C_{3z}$ eigenvalues $\exp(-2i\pi J_z/3)$, where $J_z$ is called the pseudo angular momentum\cite{zhang_chiral_2015,liu_pseudospins_2017}. TRS pairs up Bloch states of opposite pseudo angular momenta $\pm J_z$. For example, the Bloch wave functions composed of $p_{x,y}$ orbitals will be organized into $|+\uparrow\rangle$, $|-\downarrow\rangle$, $|+\downarrow\rangle$, $|-\uparrow\rangle$ states with $J_z=+3/2,-3/2, +1/2,-1/2$, respectively, where $|\pm\rangle$ refers to $p_x\pm ip_y$ and $\uparrow(\downarrow)$ denotes spin up (down). In SnH, these four states are degenerate at $\Gamma$ without SOC, and are split into $J_z=\pm3/2$ and $\pm1/2$ by SOC\cite{xu_large-gap_2013}.

\begin{figure}
\includegraphics[width=1.0\columnwidth]{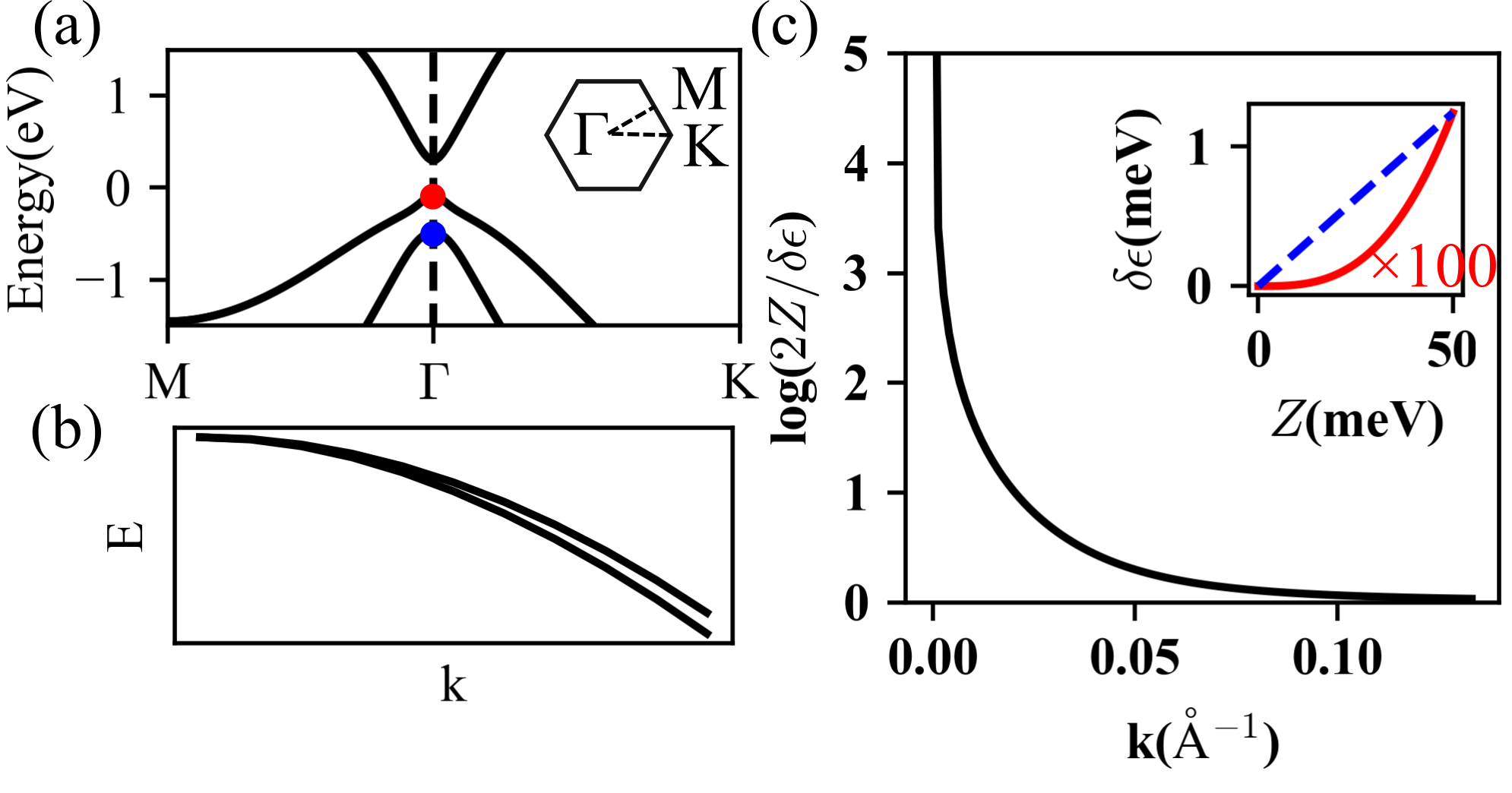}
\caption{(a) Band structure of SnH with SOC. Valence bands labeled by red (blue) dot are mainly contributed by $|+\uparrow\rangle$ and $|-\downarrow\rangle$ ( $|+\downarrow\rangle$ and $|-\uparrow\rangle$) orbitals near $\Gamma$. Inset: Brillouin zone of SnH. (b) Schematic plot of Zeeman splitting ($\delta\epsilon_+$) between $\epsilon_{++}$ and $\epsilon_{+-}$ as a function of $k$. (c) The ratio between Zeeman splitting without SOC $2Z$ ($Z=0.1$ meV in this picture) and the actual Zeeman splitting $\delta\epsilon$ as a function of $k$ in the $\Gamma$-$M$ direction for SnH. $\delta\epsilon$ is roughly independent of the polar angle of $\boldsymbol{k}$. Inset: Zeeman splitting $\delta\epsilon_+$ (red solid line, magnified by 100) and $\delta\epsilon_-$ (blue dashed line) at $\Gamma$ 
{($\boldsymbol{k}=\boldsymbol{0}$)} as a function of Zeeman field $Z$. \label{fig2}}
\end{figure}

Using the above states as basis functions, we derive a $\boldsymbol{k}\cdot\boldsymbol{p}$ Hamiltonian around $\Gamma$ to describe the valence bands of SnH. {Under the basis order $(|+\uparrow\rangle$, $|+\downarrow\rangle$, $|-\downarrow\rangle$, $|-\uparrow\rangle)$,} $C_{3z}=\exp(-i\pi\sigma_z/3)\otimes\exp(-i2\pi\tau_z/3)$,  mirror symmetry $M_x=-i\sigma_x\otimes-\tau_x$, inversion symmetry $P=\sigma_0\otimes\tau_0$, and TRS $T=-i\sigma_y\otimes\tau_xK$, where the Pauli matrices $\sigma$ and $\tau$ act on spin and orbital spaces, respectively, and $K$ denotes complex conjugation. Under the basis order $(|+\uparrow\rangle$, $|-\downarrow\rangle$, $|+\downarrow\rangle$, $|-\uparrow\rangle)$ {and using $C_{3z}\circ k_\pm = \exp(\pm i2\pi/3)k_\pm$}, the $\boldsymbol{k}\cdot \boldsymbol{p}$ Hamiltonian that commutes with the symmetry operators takes the form
\begin{equation}
H_0 = C(\boldsymbol{k})+
 \begin{pmatrix}
  M(\boldsymbol{k}) & 0 & -iA_2k_+^2 & A_1k_-^2 \\
  0 & M(\boldsymbol{k}) & A_1k_+^2 & -iA_2k_-^2 \\
  iA_2k_-^2  & A_1k_-^2  & -M(\boldsymbol{k}) & 0  \\
  A_1k_+^2 & iA_2k_+^2 & 0 & -M(\boldsymbol{k}) 
 \end{pmatrix},\label{H0}
\end{equation}
where $C(\boldsymbol{k})=C_0+C_1(k_x^2+k_y^2)$, $M(\boldsymbol{k})=M_0+M_1(k_x^2+k_y^2)$, $k_\pm=k_x\pm ik_y$, and all the coefficients are real. Here $M(\boldsymbol{k})$ refers to the SOC-induced band splitting. $A_1$ and $A_2$ terms correspond to interband mixing, which are exactly zero at $\Gamma$. To the lowest order approximation, $H^\textrm{II}_\textrm{SOC} \cong M_0 \tau_z \sigma_z$, which, if replacing the orbital index $\tau_z$ with the valley index $s_z$, looks the same as $H^\textrm{I}_\textrm{SOC}$. Hence, in analogy to the spin-valley locking induced by $H^\textrm{I}_\textrm{SOC}$,  $H^\textrm{II}_\textrm{SOC}$ induces a so-called spin-orbital locking\cite{fu_parity-breaking_2015,zhang_spin-orbital_2013}. $M_0 \tau_z$ acts as an out-of-plane Zeeman-like field, which shows opposite signs for opposing orbitals $\tau_z = \pm 1$ [Fig. \ref{fig1}(d)]. For SnH, $M_0 = 0.205$ eV {[see Fig. \ref{fig2}(a)]} {obtained by \emph{ab initio} calculations\cite{supplementary}}, equivalent to a Zeeman field $\sim3.5\times10^3$ Tesla. Electron spins are intrinsically polarized by the strong Zeeman-like field. Thus the influence of in-plane magnetic field is strongly suppressed, which will be explicitly demonstrated below. 

The two-fold degenerate bands of $H_0$ have emergent full rotational symmetry: $\epsilon_\pm=C(\boldsymbol{k})\pm\sqrt{M^2(\boldsymbol{k})+A^2k^4}$, where $A^2=A_1^2+A_2^2$ and $k^2=k_x^2+k_y^2$. A Zeeman field in the $x$ direction is introduced in the Hamiltonian as $H_{Z_\parallel} = Z_\parallel \sigma_x \tau_0$, which splits the two-fold degeneracy by breaking both $C_{3z}$ and $T$. The full Hamiltonian $H=H_0+H_{Z_\parallel}$ has eigenvalues
\begin{equation}
\epsilon_{\pm\pm}=C(\boldsymbol{k})\pm\sqrt{M^2(\boldsymbol{k})+A^2k^4+Z_\parallel^2\pm \Delta_\parallel^2(\boldsymbol{k})},
\end{equation}
where $\Delta_\parallel^2(\boldsymbol{k})=2\sqrt{4Z_\parallel^2A_2^2k_x^2k_y^2+Z_\parallel^2A_1^2k^4}$. The second $\pm$ in the subscript of $\epsilon$ denotes splitting caused by Zeeman interaction. The corresponding Bloch wave functions are denoted as $|u_{\pm\pm}\rangle$. Energy splitting caused by the Zeeman field is $\delta\epsilon_\pm \approx\Delta_\parallel^2(\boldsymbol{k})/M_0$. For small $k$, this energy splitting is much smaller than the Zeeman energy: $\delta\epsilon_\pm\ll Z_\parallel$. The band spitting caused by $H_{Z_\parallel}$ is visualized in Fig \ref{fig2}(b), where the magnitude of the Zeeman splitting is exaggerated for clearness.

According to perturbation theory, the lowest order nonvanishing term of $\delta\epsilon_\pm$ should be proportional to $Z_\parallel^2$ at $\Gamma$. However, due to the simplicity of the $\boldsymbol{k}\cdot\boldsymbol{p}$ model, the two-fold degeneracy is preserved at $\Gamma$ even for broken TRS. To break the degeneracy, we need to take orbital mixing into account. For instance, $s$ and $p_z$ have nonzero contribution to the valence bands of SnH, which are able to hybridize with $J_z = \pm 1/2$ states of $p_{x,y}$. In fact, an in-plane Zeeman interaction $\hat{\sigma}_x$ is symmetrically allowed to have the form 
\begin{equation}
 H_Z=\begin{pmatrix}
  0 & 0 & Z_1 & -iZ_3 \\
  0 & 0 & -iZ_3 & Z_1 \\
  Z_1 & iZ_3 & 0 & Z_2 \\
  iZ_3 & Z_1 & Z_2 & 0
 \end{pmatrix},\label{HZ}
\end{equation}
where $Z_2$ and $Z_3$ arise from orbital mixing.

To quantify the influence of in-plane Zeeman field, we carry out \emph{ab initio} calculations based on density functional theory and add a Zeeman interaction $Z\hat{\sigma}_x$ into the \emph{ab initio} Hamiltonian, as described in the Supplemental Material\cite{supplementary}. {The \emph{ab initio} calculations reveal that the $p_x$ and $p_y$ orbitals of Sn constitute more than 80\% of the four valence bands at $\Gamma$} and we deduce that $Z_2$ and $Z_3$ are much smaller than $Z$. Eq. (\ref{HZ}) breaks the two-fold band degeneracy at $\Gamma$. One important feature of Eq. (\ref{HZ}) is the missing of Zeeman interaction in the subspace of ($|+\uparrow\rangle$, $|-\downarrow\rangle$). Therefore, the splitting of the upper valence band should be quadratic in $Z$, while the splitting of the lower valence band should be linear in $Z$, as displayed in the inset of Fig. \ref{fig2}(c). Notice that the splitting of the lower valence band is still much smaller than $Z$. Away from $\Gamma$, the protection against in-plane Zeeman field gets weaker due to the increasingly strong interband mixing. In Fig. \ref{fig2}(c), we plot the ratio between the Zeeman splitting without SOC ($2Z$) and with SOC ($\delta\epsilon$) for the upper valence band of SnH. As expected, the ratio is extremely large near $\Gamma$ and decreases significantly away from $\Gamma$. When the Fermi level is tuned near the band edge, for instance, by chemical doping or electrical gating, novel electronic states insensitive to in-plane magnetic field will be realized.   

The suppressed Zeeman band splittings imply that superconductivity will be insensitive to in-plane magnetic field, resulting in the so-called type-II Ising superconductivity. To evaluate the upper critical field of type-II Ising superconductors, we numerically solve the self-consistent gap equation for time reversal invariant pairing, as described in the Supplemental Material\cite{supplementary}. In the limit of zero temperature, the upper critical field for SnH can be determined by:
\begin{equation}
\delta\epsilon = 2\hbar\omega_D \exp(-1/gV\nu) \approx 1.13k_BT_c,
\end{equation}
where $\hbar\omega_D$ is Debye energy, $g$ is pairing strength, $V$ is the volume of material, $\nu$ is the density of states at the Fermi energy, and $T_c$ is the critical temperature. The result indicates that the superconductivity is destroyed when the Zeeman splitting reaches $1.13k_BT_c$. As the Zeeman splitting is strongly suppressed by the SOC field [Fig. \ref{fig2}(c)], $B_{c2}$ can be greatly enhanced in type-II Ising superconductors. 

\begin{figure}
\includegraphics[width=1.0\columnwidth]{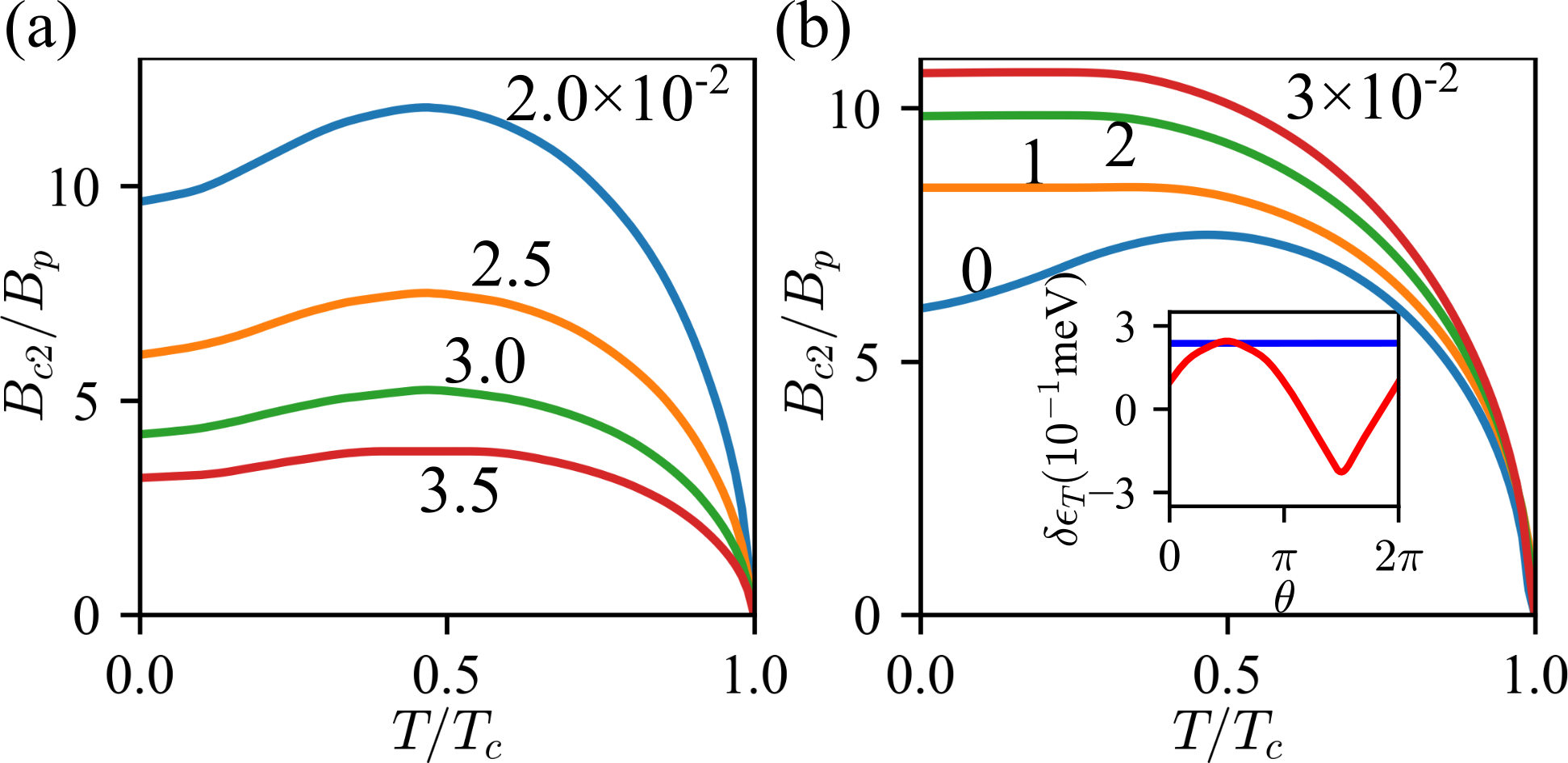}
\caption{(a) The relation between critical field $B_{c2}$, normalized by Pauli limit $B_p$, with respect to temperature $T$, normalized by critical temperature $T_c$ for $k_F=0.02/0.025/0.03/0.035$ \AA$^{-1}$. (b) The relation between critical field with respect to temperature for out-of-plane electric field $E=0.01/0.02/0.03$ V/\AA~at $k_F=0.025$ \AA$^{-1}$. Inset: $\delta\epsilon_T$ (see main text for definition) for $E=0$ (blue) and $E=0.03$ V/\AA (red) under 1 meV Zeeman field at $k_F=0.025$ \AA$^{-1}$ for varying polar angles of $k_F$. {Parameters $\hbar\omega_D=0.01$ eV\cite{peng_low_2016} and $gV\nu=0.205$ are used, giving $T_c \sim 1$ K\cite{liao_superconductivity_2018}}.\label{fig3}}
\end{figure}

$B_{c2}$ as a function of temperature is calculated for SnH with varying Fermi wave vector $k_F$ for the upper valence band. As shown in Fig. \ref{fig3}(a), indeed $B_{c2}$ can be significantly larger than the Pauli limit $B_p$, especially for small $k_F$. Shape of the $B_{c2}-T$ curve is similar to that of type-I Ising superconductors\cite{yuan_ising_2016}, implying similar physical origin yet different realization approaches. {Note that we have restricted ourselves to zero-momentum time reversal invariant pairing, and the Fulde-Ferrell-Larkin-Ovchinnikov (FFLO) state \cite{fulde1964superconductivity,larkin1965nonuniform,matsuda_fuldeferrelllarkinovchinnikov_2007} may occur near $B_{c2}$, which could further enhance $B_{c2}$.}

To simulate the influence of inversion symmetry breaking, an out-of-plane electric field is introduced in \emph{ab initio} calculations of SnH. The electric field ($\boldsymbol{E}$) will induce an in-plane SOC field $\propto \boldsymbol{k} \times \boldsymbol{E}$ by the Rashba effect, which is usually detrimental to Ising superconductivity. However, the Rashba effect is weak due to the small $\boldsymbol{k}$ here. A complicated competition between the Rashba effect, other electronic effects induced by $\boldsymbol{E}$ and the external Zeeman field thus exists. As shown in Fig. \ref{fig3}(b), the critical field increases slightly under small electric fields. To understand the result, an energy splitting is defined for the superconducting pair $|u(\boldsymbol{k})\rangle$ and $|\hat{T}u(\boldsymbol{k})\rangle$, $\delta\epsilon_T:=\langle u(\boldsymbol{k})|
\hat{H}|u(\boldsymbol{k})\rangle-\langle \hat{T}u(\boldsymbol{k})|
\hat{H}|\hat{T}u(\boldsymbol{k})\rangle$, {where $\hat{H}$ is the single particle Hamiltonian including Zeeman fields in the non-superconducting state}. As displayed in the inset of Fig. \ref{fig3}(b), the average value of $\delta\epsilon_T$ is increased by small electric field, thus giving an enhanced critical field. {Rotational symmetry breaking will generally suppress type-II Ising superconductivity, whose influence can be experimentally explored by applying strain\cite{supplementary}}. {One may further generalize the spin-orbital locking mechanism to three-dimensional crystals}, which helps to understand the unusual spin-space anisotropy of iron-based superconductors\cite{fernandes_distinguishing_2014,ma_prominent_2017,day_influence_2018}. The results imply that the mechanism of type-II Ising superconductivity works for noncentrosymmetric systems and is robust against perturbations in the presence of strong SOC.

\begin{figure}
\includegraphics[width=1.0\columnwidth]{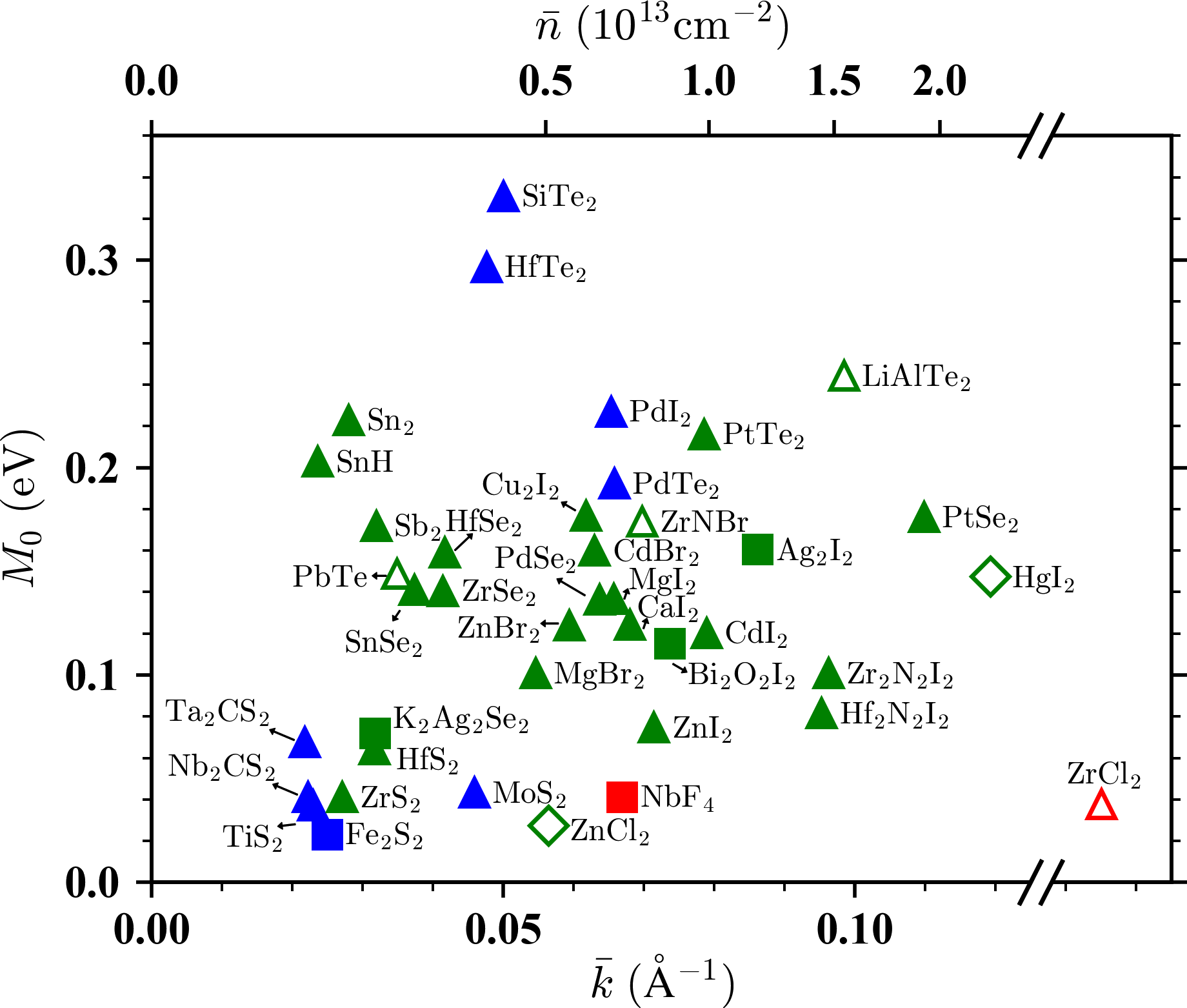}
\caption{Candidate materials of type-II Ising superconductors. Blue markers indicate the relevant bands cross the intrinsic Fermi energy, and green (red) markers indicate the relevant bands are below (above) the intrinsic Fermi energy. Solid (hollow) markers indicate inversion symmetric (asymmetric) materials. Triangular, rectangular and rhombic markers correspond to $C_3$, $C_4$, $S_4$ symmetric materials. Type-II Ising superconductivity is prominent for $k_F<\bar{k}$. The rightmost dot (ZrCl$_2$) features the coexistence of type-I and type-II Ising superconductivity with $2Z/\delta\epsilon_T$ always above 10 in our simulations. {Therefore, $\bar{k}$ is ill defined for this material}. The top axis denotes an estimation of carrier density contributed by the this pocket near $\Gamma$ with $k_F = \bar{k}$.
\label{fig4}}
\end{figure}

To identify the most promising materials for Type-II Ising superconductivity, we carry out high-throughput \emph{ab initio} calculations on 2D materials database\cite{mounet_two-dimensional_2018} and other common 2D materials. The guiding principles are follows: The candidate materials should have $C_{3/4/6}$ or other symmetries like $S_{4/6}$, so that their bands could be degenerate at high-symmetry momenta without SOC and become split by SOC. The relevant bands, if close to the Fermi level, are interesting for Ising superconductivity. Their energy splitting $\delta\epsilon_T$ caused by in-plane Zeeman field is calculated for varying $k_F$ by first-principles methods. Two quantities are used to quantify the property: $M_0$ that is the Zeeman-like field at $\boldsymbol{k} = \boldsymbol{0}$ ($2M_0$ being the spin split gap) and $\bar{k}$ defined by $2Z/\delta\epsilon_T|_{k_F = \bar{k}} = 10$. Type-II Ising superconductivity is prominent for $k_F<\bar{k}$ and for large $M_0$. About one hundred candidate materials of varying symmetries are calculated, and promising candidates are presented in Fig. \ref{fig4}. More information about their atomic configuration, band structures and Zeeman energy splittings are included in the Supplemental Material\cite{supplementary}.

{In Fig. \ref{fig4}, most candidate materials have small carrier density contributed by the Fermi pocket near $\Gamma$ ($\bar{k}\sim 0.1$\AA$^{-1}$), which is unfavorable for superconductivity either due to reduced Coulomb screening\cite{bogoljubov_new_1958,mcmillan_transition_1968,morel_calculation_1962} or disorder scattering\cite{belitz_anderson-mott_1994}. {For Coulomb repulsion depending on $\log (E_F/\hbar\omega_D)$,} we calculate $\bar{E}_F$ (energy difference between $k=\bar{k}$ and $k=0$) for the materials in Fig. \ref{fig4} and the average value is $0.036$ eV, larger than Debye energy $\sim$0.02 eV for most 2D materials\cite{peng_thermal_2016,peng_phonon_2016}. {Furthermore, the total carrier density can be enhanced by increasing film thickness. For unconventional superconductivity, the Coulomb repulsion may be also overcome by a sign-changing (as a function of momentum) order parameter.} On the other hand, the carrier mean free paths in most of current experiments are much larger than $1/\bar{k}$\cite{banszerus2016ballistic,yoon2011good}. We are also aware of many experimental discoveries of superconductors with low carrier densities \cite{fatemi_electrically_2018,wang_interface-induced_2012,coldea_key_2018,zhang_ubiquitous_2017,zhao_direct_2018}. Therefore, we believe that superconductivity is feasible in these candidate materials.}

Promising candiate materials include chalcogenides and halides in a CdI$_2$ protype structure or in a MoS$_2$ protype structure. Though only monolayer materials are calculated, their few-layer films also fit the scenario of type-II Ising superconductivity, making the study of quantum size effects interesting. Note that many candidate materials are intrinsically not superconductors. However, they might be driven into the superconducting phase, for instance, by ionic liquid gating\cite{saito_highly_2017}. Importantly, Rashba effects have minor influence on type-II Ising superconductors, which is advantageous for tuning superconductivity by gating. Noticably, in ZrCl$_2$ and ZrCl$_2$-alike materials, the SOC results in spin-orbital locking and spin-valley locking near $\Gamma$ and $K$/$K'$, respectively, making the coexistence of type-I and type-II Ising superconductivity in the same material possible.

In summary, we propose a type-II Ising superconductivity in 2D materials with SOC, which is applicable to both centrosymmetric and noncentrosymmetric materials having multiple degenerate orbitals. The type-I and type-II mechanisms can be experimentally distinguished by many features, including crystalline symmetry, spin splitting, Fermi pocket, thickness dependence and external perturbation {(such as substrate interaction and electric gating)}\cite{supplementary}. Type-II Ising superconductivity is featured by diverse candidate materials and can robustly exist when varying the film thickness and under the inversion symmetry breaking effects (e.g., substrate and electric gating)\cite{supplementary}. The finding greatly enriches the physics and materials of Ising superconductivity, which sheds new lights on future research (e.g. topological superconductivity) and device applications {(e.g. spintronics utilizing persistent spin textures\cite{schliemann_colloquium:_2017,schliemann_nonballistic_2003})}.

\emph{Note added:} Very recently, enhanced $B_{c2}$ was found experimentally in few-layer stanene\cite{falson_type-ii_2019} and 1T-PdTe$_2$\cite{liu_quantum_2019}, providing strong support to the predicted type-II Ising superconductivity. 

\begin{acknowledgments}
Chong Wang and Biao Lian contributed equally to this work. We thank Naoto Nagaosa and Jun He for helpful discussions. This work was supported by the Basic Science Center Project of NSFC (Grant No. 51788104), the Ministry of Science and Technology of China (Grants No. 2016YFA0301001, No. 2018YFA0307100, No. 2017YFA0304600, No. 2017YFA0302902 and No. 2018YFA0305603), the National Natural Science Foundation of China (Grants No. 11674188, No. 11790311 and No. 11874035) and the Beijing Advanced Innovation Center for Materials Genome Engineering. B.L. acknowledges the support of Princeton Center for Theoretical Science at Princeton University.
\end{acknowledgments}

\bibliography{main}

\end{document}